\documentstyle[aps,preprint]{revtex}
\begin{document}
\draft
\title{Mirror matter}
\author{ A.Yu.Ignatiev and
R.R.Volkas}
\address{School of Physics, Research Centre for High Energy
Physics, University of Melbourne, Victoria 3010,
Australia}
\maketitle
\begin{abstract}
One of the deepest unsolved puzzles of subatomic physics is why Nature
prefers the left particles to the right ones. Mirror matter is an attempt
to understand this mystery by assuming the existence of a "parallel"
world where this preference is exactly opposite. Thus in the Universe
consisting of the ordinary and the mirror matter the  symmetry between
the left and right is completely restored. Mirror matter is constrained
to interact with us only very weakly. Still, its existence can be
inferred by using experimental evidence such as the observation of
astrophysical objects related to the dark matter (MACHO), neutrino
physics and other sources. This talk will focus on several key aspects of
mirror matter physics including the possible existence of mirror matter
inside the Earth and the suggestion that the recently observed "isolated"
planets may  in fact be orbiting around mirror stars.
\end{abstract}
1.A variety of observations strongly suggest the presence of a
significant amount of dark matter (DM) in the universe. Galactic
rotation curves and cluster dynamics cannot be explained using standard
Newtonian gravity unless non-luminous but gravitating matter exists.
Arguments from Big Bang Nucleosynthesis (BBN)(Recent results \cite{boom}
based on the measurements of the Cosmic Microwave Background anisotropy
are in agreement with the BBN constraint on the baryon density)
 and theories of large scale
structure formation disfavour the simple possibility that all of the DM
consists of ordinary baryons. Candidates for the required exotic
component in the DM abound: WIMPS, axions and mirror matter are
examples. The observation of microlensing events from the Small and
Large Magellanic Clouds is consistent with the existence of Massive
Compact Halo Objects (MACHOs) in the halo of the Milky Way \cite{macho}.
The inferred average mass is about $0.5 M_{\odot}$, where $M_{\odot}$ is
the mass of our sun. The most reasonable conventional identification
sees MACHOs as white dwarfs, although there are several strong arguments
against this \cite{freese}. For example, the heavy elements that would
have been produced by their progenitors are not in evidence
\cite{freese}. This argues against the conventional white dwarf
scenario, and in favour of exotic compact objects. In summary, there is
strong evidence for exotic DM which is capable of forming compact
stellar mass objects. Mirror matter \cite{v3,v4,v6,v7} is an interesting
candidate for some of the required exotic DM \cite{v6,v6a}. It can be
independently motivated by the desire to see the full Poincar\'{e}
Group, including improper transformations (parity and time reversal), as
an exact symmetry group of nature. The basic postulate is that every
ordinary particle (lepton, quark, photon, etc.) is related by an
improper Lorentz transformation with an opposite parity partner (mirror
lepton, mirror quark, mirror photon, etc.) of the same mass. Both
material particles (leptons and quarks) and force carriers (photons,
gluons, $W$ and $Z$ bosons) are doubled. Mirror matter interacts with
itself via mirror weak, electromagnetic and strong interactions which
have the same form and strength as their ordinary counterparts (except
that mirror weak interactions couple to the opposite chirality). Because
ordinary matter is known to clump into compact objects such as stars and
planets, mirror matter will also form compact mirror stars and mirror
planets. Since mirror matter does not feel ordinary electromagnetism, it
will be dark. Gravitation, by contrast, is common to both sectors.
Mirror matter therefore has the correct qualitative features: it is
dark, it clumps, and it gravitates. It has been speculated that MACHOs
might be mirror stars \cite{f1}, and the observed extrasolar planets
\cite{extrasolar} might be composed of mirror matter \cite{f2}.

2.An important question that arises naturally is whether or not the
existence of mirror particles can lead to other observable consequences.
In particular, it is essential to find constraints on the possible
concentration of mirror particles in the Earth. Two main approaches to
our problem are possible. First, one can trace the fate of the mirror
particles starting from the early Universe epoch through the structure
formation periods (galaxies, solar system and finally the Earth).
Second, we can use geophysical data to get a more direct limit on the
concentration of mirror matter in the Earth regardless of possible
cosmological bounds. It has been suggested that considerations based on
the structure formation theory disfavour a significant presence of
mirror matter in the Earth \cite{v6}. However, as our knowledge of the
structure formation with mirror matter is still incomplete, it is
important to develop a geophysical approach as an independent,
complementary tool of analysis exploiting the wide and rich variety of
observational data accumulated in the Earth sciences. This approach is
applicable not only to the specific mirror matter model, but also to any
other theory predicting the existence of a new world of particles which
couples to the ordinary matter only through gravitational interaction.
An example is the shadow matter characteristic of superstring theories.
For a detailed investigation of geophysical constraints on the possible
admixture of mirror matter inside the Earth a method has been developed
\cite{iv} based on the Preliminary Reference Earth Model \cite{5,6}
---the ``Standard Model'' of the Earth which describes its
internal structure derived from the geophysical data in a systematic and
self-consistent manner. If the density of the mirror matter is given,
our method allows one to compute changes in various quantities
characterising the Earth (such as its mass, moment of inertia,
frequencies of its normal modes etc.). Comparing the computed and
observed values of these characteristics, we can obtain for the first
time the direct upper bounds on the possible concentration of the mirror
matter in the Earth. In terms of the ratio of the mirror mass to the
Earth mass these upper bounds range from $4\times 10^{-4}$ to
$4\times 10^{-3}$  depending on the radius of the mirror matter ball.
 Furthermore, it has been possible to analyze various manifestations
of mirror matter through the variations of the gravity acceleration on
the Earth surface. These variations could arise as a result
of an off-centre shift of the mirror matter due to several possible
mechanisms such as lunar and solar tidal forces, meteorite impacts
 and earthquakes.
Our estimates have shown that variations caused by these mechanisms seem
too small to be observed. In \cite{iv} our analysis was based on a
standard premise that mirror matter interacts with ordinary matter only
gravitationally. Note that mirror matter can also couple to ordinary
matter through photon---mirror-photon kinetic mixing \cite{v4,new}. Some
implications of this interaction have recently been discussed in
Ref.\cite{fiv}. The presence of such mixing can open more possibilities
for non-gravitational ways of observing mirror DM. Another way in which
ordinary matter can interact with mirror matter is through
Higgs---mirror-Higgs mixing\cite{v4} and some implications of this have
been recently analysed in Ref.\cite{iv1}, but we have not relied on
these assumptions about the mirror matter properties. Therefore our
results are valid for other types of hypothetical matter coupled to
ordinary matter by gravitation only; an example is shadow matter
introduced in string theories. On the other hand, the use of equation of
state and other macroscopic characteristics of mirror matter could lead
to more severe constraints on the mirror mass inside the Earth.

3.Zapatero Osorio et al.\ \cite{zo} have recently presented strong
evidence for the existence of ``isolated planetary mass objects''
in the $\sigma$ Orionis star cluster. These objects are more
massive than Jupiter $M_{J}$, but not as massive as brown dwarfs
($\sim 5 - 15 M_J$ although there is some model dependence in the
mass determination\cite{zo}). They appear to be gas giant planets
which do not seem to be associated with any visible star.
Given that the
$\sigma$ Orionis cluster is estimated to be between 1 million and
5 million years old, the formation of these ``isolated planets''
must have occured within this time scale. Zapatero Osorio et al.\
argue that these findings pose a challenge to conventional
theories of planet formation because standard theories of
substellar body formation (as well as new theories inspired by
previous claims of isolated planet discovery), are unable to
explain the existence of numerous isolated planetary mass objects
down to masses $\sim$ few $M_J$. See Ref.\cite{zo,zo1,lada} and references
therein for further discussion. It is possible therefore that
non-standard particle physics may be required to understand their
origin.
We have speculated \cite{fiv1} that rather than being isolated, these
ordinary matter planets actually orbit invisible mirror stars.
It should be possible to test this idea by searching for a
periodic Doppler shift in spectral lines emanating from these
planets.
Suppose that a given planet is in a circular orbit of radius $r$ around
a mirror star of mass $M$. Let $I$ be the inclination of the plane of
the orbit relative to the normal direction defined by the Earth - mirror
star line. Then we obtain
\begin{equation}
\frac{\Delta \lambda}{\lambda} \simeq 10^{-3}
\sqrt{\frac{M}{M_{\odot}}} \sqrt{\frac{0.04\ A.U.}{r}} \sin I
\end{equation}
as the level of spectral resolution required, where $\lambda$ is
wavelength, $\Delta \lambda$ is the difference between the peak and
trough of the periodic Doppler modulation of $\lambda$. Note that this
is a few orders of magnitude larger than the Doppler shifts observed in
extrasolar planet detection. However it is certainly true that the
isolated planets are much fainter sources of light than the stars whose
Doppler shifts have been measured, so such a measurement may not be
completely straightforward. Yet it is worth noting that for the case of
close orbiting ordinary planets where $r \sim 0.04\ A. U.$ (analogous to
the close-in extra solar planets), the Doppler shift is quite large
($\sim 10^{-3}$) with a period of only a few days which should make this
interesting region of parameter space relatively easy to test. Indeed,
Zapatero Osorio et al.\ \cite{zo} have taken optical and near infrared
low resolution spectra of three young isolated planet candidates (S Ori
52, S Ori 56, and S Ori 47). They have obtained absorption lines (at
wavelengths $\sim 900$ nm), however their resolution was 1.9 nm\cite{zo}
which is just below that needed to test our hypothesis. The higher
resolution required has been achieved in the case of brown
dwarfs\cite{brown} so we anticipate that it should be possible to test
our hypothesis in the near future. One would also expect some ordinary
matter to have accumulated in the centre of the mirror star. It is
possible, but not inevitable, that this ordinary matter also observably
radiates. If so, one would expect this radiation to experience a much
smaller Doppler modulation compared to that from the planet. Because the
planet and mirror star would not be spatially resolved, one
observational signature would be that some of the spectral lines are
modulated (those from the planet), while a different set are not (those
from the ordinary matter pollutants in the mirror star). If the mirror
star is invisible but opaque, then one would expect to see periodic
planetary eclipses for some of these systems (those with $\sin I \simeq
1$). The eclipses should of course occur once per Doppler cycle, around
one of the points of zero Doppler shift within a cycle. Obviously, such
eclipses (along with the information provided by Doppler shift
measurements) will be useful in distinguishing a mirror star from
alternatives such as faint white dwarfs or neutron stars. However, it
should be mentioned that standard objects such as white dwarfs and
neutron stars are  unlikely candidates, because the age of the $\sigma$
Orionis cluser is estimated to be only 1 million to 5 million years old,
while white dwarfs  are typically billions of years old and neutron
stars are generally tens to hundreds million year old. Before
concluding, we would like to point out an intriguing systematic in both
the extrasolar planet and the Zapatero Osorio et al.\ data that may
argue in favour of the mirror matter hypothesis: one might expect the
number of hybrid systems to grow up as a function of the disparity
between the components and indeed the
Zapatero Osorio et al.\ objects also increase in number with decreasing
mass \cite{zo1}.
Of course if the ``isolated
planets'' do orbit mirror stars then this suggests that the star forming
region near $\sigma$ Orionis could also be a region of mirror star
formation. This is certainly possible and was already envisaged many
years ago by Khlopov et al.\cite{kh} where they argued that large
molecular clouds (made of ordinary matter) could merge with large mirror
molecular clouds in which case the formation of mixed systems (i.e.
containing both ordinary and mirror matter) is enhanced.

4.In conclusion, we have outlined the results of a detailed investigation
of geophysical constraints
 on the possible
admixture of mirror matter inside the Earth. On the basis of the
Preliminary Reference Earth Model (PREM)---the ``Standard Model'' of the
Earth's interior---we have developed a method which allows one to
compute changes in various quantities characterising the Earth (mass,
moment of inertia, normal mode frequencies etc.) due to the presence of
mirror matter. As a result we have been able to obtain for the first
time the direct upper bounds on the possible concentration of the mirror
matter in the Earth. In terms of the ratio of the mirror mass to the
Earth mass a conservative upper bound is $4\times 10^{-3}$.
 Also, it is possible to analyse various mechanisms
(such as lunar and solar tidal forces, meteorite impacts and
earthquakes) of exciting mirror matter oscillations around the Earth
centre. Such oscillations could manifest themselves through global
variations of the gravitational acceleration at the Earth's surface. Our
results show that such variations are too small to be observed. Our
conclusions are valid for other types of hypothetical matter coupled to
ordinary matter by gravitation only (e.g. the shadow matter of
superstring theories).
 Also, we have described the proposal that the recently observed
 ``isolated'' planetary mass
objects might actually be planets orbiting invisible mirror stars. This
idea can be tested by searching for a Doppler modulation at the level of
$10^{-3}-10^{-4}$ in amplitude.
The authors are grateful to R.Foot, G.C.Joshi and B.H.J.McKellar for
interesting discussions. This work was supported in part by the
Australian Research Council.

\end{document}